\pdfoutput=1 
\documentclass[%
 aip,
 rsi,
 amsmath,amssymb,
reprint,%
]{revtex4-1}

\usepackage{graphicx}
\usepackage{dcolumn}
\usepackage{bm}
\usepackage{tikz}
\usepackage[american, betterproportions, oldvoltagedirection]{circuitikz} 
\usetikzlibrary{shapes, calc, arrows, positioning, external}

\usepackage{adjustbox}

\usepackage[utf8]{inputenc}
\usepackage[T1]{fontenc}
\usepackage{mathptmx}

\newcommand{\comment}[2]{\hspace{0in}#2}

\begin{document}

\preprint{AIP/123-QED}
\title[]{Current-limiting amplifier for high speed measurement of resistive switching data}

\author{T. Hennen}
 \email{tyler@hennen.us}
 \affiliation{IWE II, RWTH Aachen University, 52074 Aachen, Germany}
 \author{E. Wichmann}
 \affiliation{IWE II, RWTH Aachen University, 52074 Aachen, Germany}
\author{A. Elias}
  \affiliation{Western Digital San Jose Research Center, 5601 Great Oaks Pkwy, San Jose, CA 95119}
\author{J. Lille}
  \affiliation{Western Digital San Jose Research Center, 5601 Great Oaks Pkwy, San Jose, CA 95119}
\author{O. Mosendz}
  \affiliation{Western Digital San Jose Research Center, 5601 Great Oaks Pkwy, San Jose, CA 95119}
\author{R. Waser}
 \affiliation{IWE II, RWTH Aachen University, 52074 Aachen, Germany}
\author{D. J. Wouters}
 \affiliation{IWE II, RWTH Aachen University, 52074 Aachen, Germany}
\author{D. Bedau}
  \email{daniel.bedau@wdc.com}
  \affiliation{Western Digital San Jose Research Center, 5601 Great Oaks Pkwy, San Jose, CA 95119}

\date{\today}

\begin{abstract}
Resistive switching devices, important for emerging memory and neuromorphic
applications, face significant challenges related to control of delicate
filamentary states in the oxide material. As a device switches, its rapid
conductivity change is involved in a positive feedback process that would lead
to runaway destruction of the cell without current, voltage, or energy
limitation. Typically, cells are directly patterned on MOS transistors to limit
the current, but this approach is very restrictive as the necessary integration
limits the materials available as well as the fabrication cycle time. In this
article we propose an external circuit to cycle resistive memory cells,
capturing the full transfer curves while driving the cells in such a way to
suppress runaway transitions. Using this circuit, we demonstrate the acquisition
of $10^5$ $I,V$ loops per second without the use of on-wafer current limiting
transistors. This setup brings voltage sweeping measurements to a relevant
timescale for applications, and enables many new experimental possibilities for
device evaluation in a statistical context.
\end{abstract}

\maketitle
\section{\label{sec:intro}Introduction}


Today, much effort is focused on employing emerging
materials and physical mechanisms for the purpose of data storage and
computation\cite{wouters_phase-change_2015, yu_emerging_2016, burr_neuromorphic_2017, sangwan_neuromorphic_2020, ma_non-volatile_2019}. Several schemes make use of Resistive Switching (RS), which refers to a large
class of related phenomena wherein the resistance of a two-terminal device can
be controlled via electrical stimuli \cite{ielmini_resistive_2015}. These
effects can be used, as in Resistive Random Access Memory (RRAM), to store bits
as non-volatile resistance states. Resistive switches can be fabricated using
wide variety of CMOS-compatible materials, and are highly attractive due to
their simple device structure, high speed, scalability, and potential for 3D
integration as required by next generation memory and computing architectures.

A central challenge for RRAM is the intrinsically stochastic nature of the RS
process, which leads to large variability in the programmed resistance states
and switching parameters \cite{chen_variability_2011, fantini_intrinsic_2013}.
Achieving an acceptable level of control over the switching process will require
an in-depth understanding of the statistical processes at play, as well as an
optimization of active material together with the control circuitry. For this
purpose, it is necessary to drive memory cells through a statistically
significant number switching cycles, and to rapidly test different materials and
modes of operation on a wafer probing system.

\begin{figure}[h]
\includegraphics[width=\columnwidth]{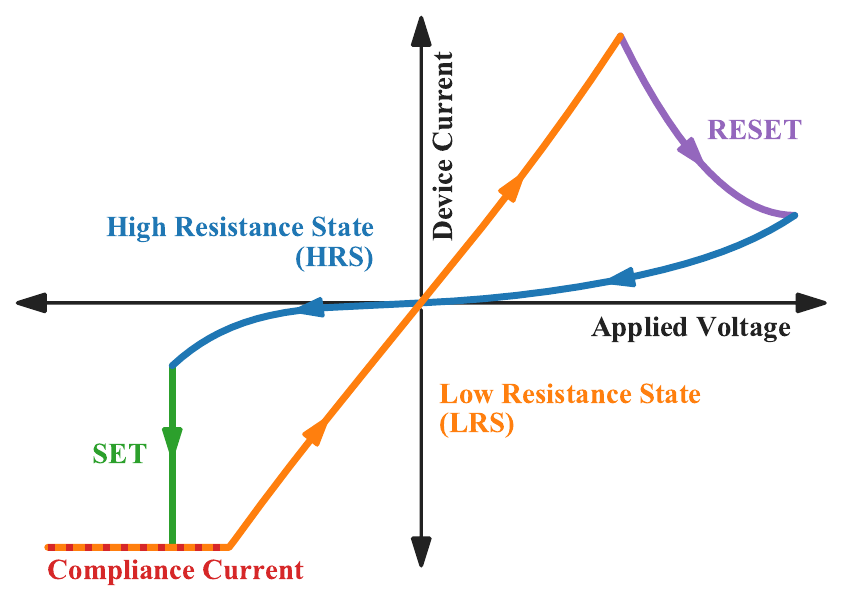}
\caption{\label{fig:ivloop_diagram} Schematic diagram of a single cycle $I,V$
loop measurement. Such loops show statistical variation device-to-device and
cycle-to-cycle, and important switching parameters may be extracted from their
measurement.
}

\end{figure}

RRAM is commonly benchmarked by direct application of square voltage pulse
sequences, but one of the shortcomings of this approach is that
only the resulting resistance states are typically recorded, while the dynamics
of the conductance changes in the material are very often left unmeasured.
Quasistatic $I,V$ loops are an alternative measurement where switching is
induced by an applied voltage that is continuously ramped at low speed
($\sim$1~V/s) between positive and negative values. The current resulting from
these sweeps is sampled and plotted against the applied voltage as shown in
Fig.~\ref{fig:ivloop_diagram}. Such $I,V$ loops are relatively rich in
information, and important parameters such as the resistance non-linearity,
voltage/current switching thresholds, and details of the transition behavior can
be extracted. However, the low speed of the measurement puts excessive
electrical stress on the device and makes experiments involving more than a few
hundred switching cycles impractical.

\begin{figure}[h]
\includegraphics[width=\columnwidth]{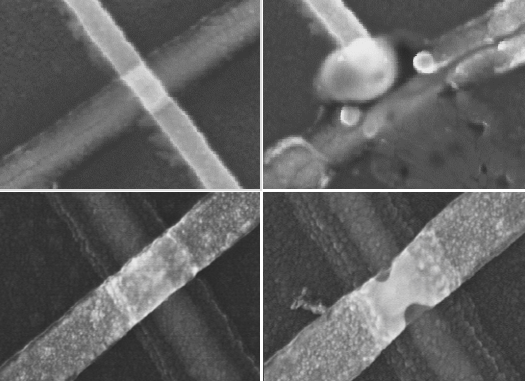}
\caption{\label{fig:destruction}
  Crossbar structures (100~nm) before (left) and after (right) being subjected to a
  current overshoot induced by an SMU with current compliance (top), and
  capacitive discharge of a 20~cm coaxial cable placed after a current limiting
  series resistance (bottom). Although the cells are visibly destroyed, both
  nevertheless continued \comment{proceeded?} to show measurable RS behavior, as
  a conducting path still existed through what remained of the oxide material.
  }
\end{figure}

While negative resistance transitions in RS materials can occur on timescales
below 1~ns\cite{torrezan_sub-nanosecond_2011, menzel_ultimate_2019,
von_witzleben_study_2020}, the nanoscale material volumes involved cannot
normally survive prolonged exposure to the voltage required to initiate the
transition, as the current density quickly reaches levels that cause
irreversible thermal damage\cite{meng_temperature_2020}. Thus, $I,V$ loop
measurements are only possible in the context of a feedback mechanism to prevent
runaway destruction of the RS device. Externally implemented current limiting
such as the current compliance function of commercial Source Measure Units
(SMUs) and Semiconductor Paramater Analyzers (SPAs) are known to cause
\comment{unacceptably} large overshoots that can lead to catastrophic damage to
cells\cite{lu_elimination_2012, tirano_accurate_2011} (Fig.
\ref{fig:destruction}) and can otherwise strongly influence the
measurements\cite{kinoshita_reduction_2008, ambrogio_analytical_2016}.
Patterning RS devices directly on MOS transistors provides superior current
limiting, but the required integration limits the materials available and
necessitates long fabrication cycle times\cite{kinoshita_reduction_2008,
nardi_control_2011, nguyen_advanced_2015}. A simpler approach from the point of
view of fabrication is to integrate fixed resistors in series with the devices
\cite{fantini_intrinsic_2012, fan_direct_2015}. However, the large linear
feedback introduced by this relatively inflexible method significantly affects
the switching behavior \cite{hardtdegen_improved_2018, gonzalez_current_2020},
and can push the operating voltage outside of the practical range.

The circuit design reported in this work represents a new way to characterize RS
devices. It can be used to suppress current overshoots and collect very large
volumes of $I,V$ sweeping characteristics without the requirement of CMOS
integration. We demonstrate collection of $10^5$ switching cycles per second,
which is highly useful for studying the stochastic nature of switching
processes.

\section{\label{sec:circuit}External current limiting amplifier}
\subsection{\label{sec:design_principles}Design principles}

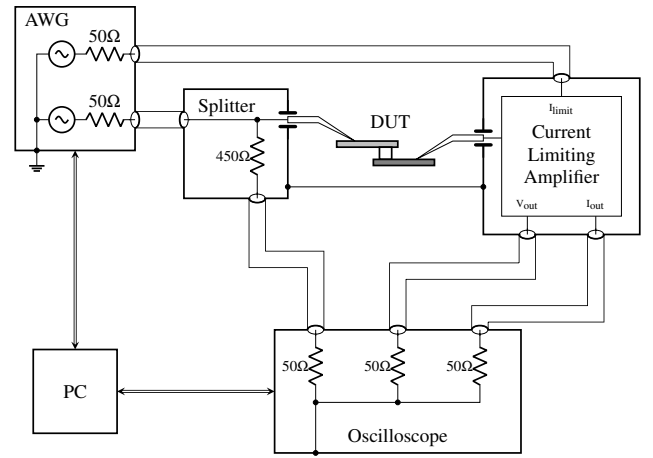
\begin{figure}[h!]
\setlength{\fboxrule}{0pt}
\setlength{\fboxrule}{0pt}
\centering
\fbox{
\maxsizebox{\columnwidth}{!}{\newcommand{\coaxDown}[1]{\draw[fill=white](#1)circle(.13cm and .07cm) --++(0,.1);\draw[thick]($(#1)+(.135,0)$)arc(0:-180:.135cm and .07cm);}
\newcommand{\coaxUp}[1]{\draw[fill=white](#1)circle(.13cm and .07cm) --++(0,-.1);\draw[thick]($(#1)+(.135,0)$)arc(0:180:.135cm and .07cm);}
\newcommand{\coaxRight}[1]{\draw[fill=white](#1)circle(.07cm and .13cm);\draw (#1) --++(-.1,0);\draw[thick]($(#1)+(0,-.135)$)arc(-90:90:.07cm and .135cm);}
\newcommand{\coaxLeft}[1]{\filldraw[fill=white](#1)circle(.07cm and .13cm);\draw (#1) --++(.1,0);\draw[thick]($(#1)+(0,.135)$)arc(90:270:.07cm and .135cm);}

\begin{tikzpicture}
\definecolor{amber}{RGB}{230,153,0}		
\def \coax{.255cm};						
\def \coaxCol{white};					

\node (TE) at (4,5) {};
\filldraw[fill=lightgray, thick](TE)++(-.3,0) rectangle ++(1,-.1) node(celltop){};
\filldraw[fill=white, thick] (celltop) ++(-.1,0) rectangle ++(-.2,-.2) node(cellbot){};
\filldraw[fill=gray, thick] (cellbot) ++(-.1,0) rectangle ++(1,-.1) ++(-.3,.1)node(BE){};
\node[above=.2] at (celltop.west){DUT};

\draw (TE.center) --++(-.6,.3)node(probeheight){} --++(-.5,0) --++(0,.05) node(TEprobe){} --++(0,.05) --++(.5,0) --(TE.center);

\draw (BE.center) --++(.6,.3) --++(.5,0) --++(0,.05) node(BEprobe){} --++(0,.05) --++(-.5,0) --(BE.center);

\node[qfpchip, num pins=12,anchor=bpin 7, hide numbers, no topmark, external pins width=0,xscale=1, yscale=1.2] (awg) at ($(TEprobe) + (-2.5,0)$){};
\node at ($(awg.bpin 12)+(.1,-.2)$) {AWG};
\node (awgCC) at ($(awg.bpin 9) !.4! (awg.bpin 8)$){};
\node (awgSplitter) at (awg.bpin 7){};

\draw(awgSplitter.center)--++(-.2,0) node[resistorshape, anchor=right, scale=.7](R){\large{$50\Omega$}}  
(R.left)--++(-.2,0)node[vsourcesinshape, rotate=90, scale=.5, anchor=south](v){};
\draw(v.north) -- ++(-.2,0)node[circ,scale=.7](Vawg){};

\draw(awgCC.center)--++(-.2,0) node[resistorshape, anchor=right, scale=.7](R){\large{$50\Omega$}}  
(R.left)--++(-.2,0) node[vsourcesinshape, rotate=90, scale=.5, anchor=south](v){};
\draw(v.north)-|(Vawg) -- (Vawg |- awg.south) node[circ,scale=.7]{} node[ground]{};


\node[dipchip, rotate=90, num pins=18, hide numbers, no topmark, external pins width=0,yscale=.8, xscale=1.2](osc) at ($(celltop) + (0,-4)$){};
\node[above] at (osc.bpin 5){Oscilloscope};
\node(oscSplitter)at(osc.bpin 17){};
\node(oscV)at(osc.bpin 14){};
\node(oscI)at(osc.bpin 11){};

\draw (oscSplitter.center) --++(0,-.3) node[resistorshape, anchor=right, rotate=90, scale=.7](R){}node[left] at (R){\scriptsize{$50\Omega$}} 
(R.left) -- ++(0,-.3)node[circ, scale=.7](gnd1){} -- (osc.bpin 2) node[circ,scale=.7]{};
\draw (oscV.center) --++(0,-.3) node[resistorshape, anchor=right, rotate=90, scale=.7](R){}node[left] at (R){\scriptsize{$50\Omega$}}
(R.left){} -- (gnd1-|R.left) node[circ, scale=.7](gnd2){} -- (gnd1);
\draw (oscI.center) --++(0,-.3) node[resistorshape, anchor=right, rotate=90, scale=.7](R){} node[left] at (R){\scriptsize{$50\Omega$}} 
(R.left) |- (gnd2);

\ctikzset{multipoles/thickness=1}
\node[qfpchip, num pins=12, hide numbers, no topmark, external pins width=0, anchor=bpin 2]
(compliance) at ($(BEprobe.center)+(.3,-.3)$){\begin{tabular}{c}Current \\ Limiting \\ Amplifier \end{tabular}};				

\draw (BEprobe.center)--(compliance.bpin 2 |- BEprobe.center) ;
\node[above]at (compliance.bpin 4){\tiny{$\mathrm{V_{out}}$}};												
\node[above]at (compliance.bpin 6){\tiny{$\mathrm{I_{out}}$}};												
\node[below]at (compliance.bpin 11){\tiny{$\mathrm{I_{limit}}$}};											

\draw (compliance.bpin 4)--++(0,-.3)node(voltageMeas){};
\draw[double=\coaxCol, double distance = \coax](voltageMeas.center) |-++(-1,-.6)-|(oscV.center);
\coaxDown{voltageMeas}
\coaxUp{oscV.center}

\draw(compliance.bpin 6)--++(0,-.3)node(currentMeas){};
\draw[double=\coaxCol, double distance = \coax](currentMeas.center) |-++(-1,-1.3)-|(oscI.center);
\coaxDown{currentMeas};
\coaxUp{oscI.center}

\draw (compliance.bpin 11)--++(0,.3)node(complianceAWG){};
\draw[double=\coaxCol, double distance = \coax](awgCC.center)-|(complianceAWG.center) ;
\coaxUp{complianceAWG};
\coaxRight{awgCC.center};

\draw[thick](BEprobe.north)node[rground,anchor=south]{}
(BEprobe.north)|-(complianceAWG.west)
(complianceAWG.east)-|($(compliance.pin 7) + (0.3,0)$)|-(currentMeas.east)
(currentMeas.west)--(voltageMeas.east)
(voltageMeas.west)-|(BEprobe.south)node[rground,anchor=south, thick,rotate=180]{};

\draw (TEprobe.center)--++(-.5,0)node[circ, scale=.7](splitter){};
\draw (splitter)--++(0,-.3) node[resistorshape, anchor=right, rotate=90, scale=.7](R){}
(R.left)--++(0,-.4)node(splitterOsc){} ;
\node[left] at (R){\scriptsize{$450\Omega$}};

\draw[double=\coaxCol, double distance = \coax](splitterOsc.center)  |-++(.5,-1)-|(oscSplitter.center);
\coaxDown{splitterOsc}
\coaxUp{oscSplitter.center}
\draw(splitter.center)--++(-1.2,0)node(splitterAWG){};
\draw[double=\coaxCol, double distance = \coax](splitterAWG.center)--(awgSplitter.center);

\coaxLeft{splitterAWG.center};
\coaxRight{awgSplitter.center};

\node[] at ($(splitter)+(-.5,.2)$){Splitter};

\draw[thick](TEprobe.north)node[rground,anchor=south]{}
(TEprobe.north)|-($(splitterAWG) + (0,.5)$)-|(splitterAWG)
(splitterAWG.south)|-(splitterOsc.west)
(splitterOsc.east)-|(TEprobe.south)node[rground,anchor=south, thick,rotate=180]{};

\node[circ, scale=.7](splitterGnd)at ($(TEprobe)+(0,.-1.1)$){};
\node[circ, scale=.7](complianceGnd)at (splitterGnd -| BEprobe){};
\draw(splitterGnd)--(complianceGnd);


\node[qfpchip, num pins=12, no topmark, external pins width=0, hide numbers, scale=.7, thick](pc) at (awg.bpin 5 |- osc) {};
\draw[<->,double,>=stealth](pc.bpin 11) -- (awg.bpin 5);
\draw[<->,double,>=stealth](pc.bpin 8)--(osc.north);
\node at (pc){PC};

\end{tikzpicture}}
      }
      \caption{\label{fig:setup_schematic}
      Schematic of a measurement setup using the current limiting amplifier
      circuit. A two channel arbitrary waveform generator (AWG) applies a driving signal to the DUT as well as
      a signal to control the value of the forward current limit. An oscilloscope
      measures simultaneous samples of the voltage at both electrodes, as well as the
      device current. A jumper connects the ground planes of the left and right probes
      to reduce interference and inductance in the signal path.
    }\end{figure}

For the purpose of rapidly testing devices with minimal nano-fabrication
overhead, compatibility with isolated two-terminal
 structures is necessary and should be
provided by an external Current Limiting Amplifier (CLA) circuit placed in
series with the Device Under Test (DUT) in a setup similar to that shown in
Fig.~\ref{fig:setup_schematic}. When the series combination is driven by a
voltage waveform, the circuit should provide a variable current limit in the
approximate range 10~$\mu$A$-$1~mA in the forward polarity (SET direction).
Because of the self-limiting nature of the RESET process under voltage control,
current should flow through the circuit unimpeded in the reverse polarity (RESET
direction).

\begin{figure}[h!]
\includegraphics[width=\columnwidth]{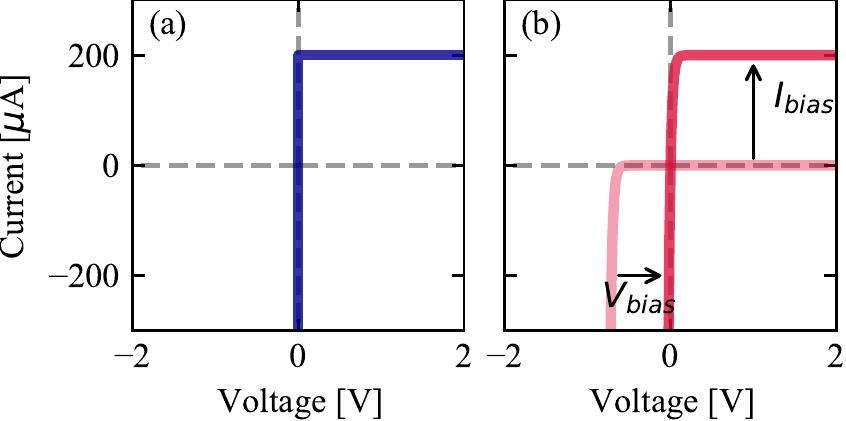}
\caption{\label{fig:limiting_characteristic}
  The current limiting $I,V$
  characteristic intended to be placed in series with the DUT. In the ideal case
  (a), the differential resistance is zero below the adjustable current limit
  (here 200~$\mu$A), and infinite above. An approximation (b) can be realized
  using a common-base amplifier with voltage and current bias.
}
\end{figure}

To avoid any influence of the circuit on the switching process before
the current limit is reached, the circuit should present a negligible impedance
for all currents below the limit. Only once the DUT current reaches the limit,
the circuit should rapidly transition into a current source behavior to
terminate the runaway switching process. In other words, the circuit should
ideally present a frequency independent $I,V$ characteristic as shown in
Fig.~\ref{fig:limiting_characteristic}(a) in series with the device. The circuit
should be highly stable for a variety of loads, and its design should be as
simple as possible in order to easily distinguish the role of the DUT in
measurements of the overall electrical response.

Crucially, any overshoot above the current limit following a SET transition
should be suppressed as much as possible. Because such overshoots are caused by
the stray capacitance at the terminal of the current limiting
circuit, this capacitance is considered a critical design parameter to be
minimized. It is therefore not an option to connect the CLA to the DUT over a
length of coaxial cable, as this would present an effective capacitance of
100~pF/m. To reduce this capacitance, the probing circuit needs to
be mounted as close as possible to the DUT, and a short unshielded probe needle
should be mounted directly to its circuit board.

The measurement setup should allow application of voltage signals of at least
1~MHz to the device, and a low-noise current output should have a resolution
below 1~$\mu$A and should be able to register fast rise times of switching
events below 100~ns. The bandwidth of signal application and current measurement
should not depend strongly on the resistance state of the DUT, nor on the
current limit used. External commercial equipment should be used to generate and
sample voltage waveforms, where an important requirement is a large enough
memory capacity to capture hundreds to thousands of $I,V$ datapoints each
switching cycle for $10^5-10^6$ cycles per measurement shot.

\subsection{\label{sec:implementation}Implementation}

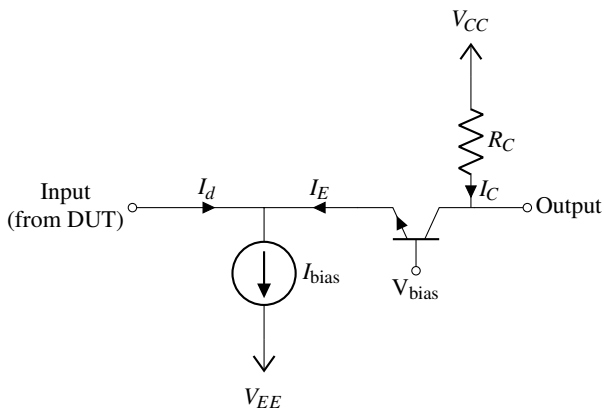
\begin{figure}[h!]
\setlength{\fboxrule}{0pt}%
\setlength{\fboxrule}{0pt}%
\centering
\fbox{
\begin{circuitikz} \draw
(0,0) node[anchor=east, align=center]{Input\\(from DUT)}
      to[short, o-, i=$I_d$] (2,0)
(1.75,0) to[isource, -, l=$I_{\text{bias}}$] (1.75,-1.75) node[vee]{$V_{EE}$}
(2,0) to[short, i<=$I_E$] (3,0)
(3,0) node[npn, xscale=-1, rotate=90, anchor=E](npn1){}
      (npn1.B) to[short, -o] (npn1.B) node[anchor=north, align=center]{V$_{\text{bias}}$} 
      (4.5, 0) to[R, l_=$R_C$, -, i<_=$I_C$] (4.5, 1.75) node[vcc]{$V_{CC}$}
      (npn1.C) to[short, -o] (5.25,0) node[anchor=west]{Output}
;\end{circuitikz}
}
\caption{\label{fig:simplified_circuit}
  A simplified diagram of a circuit implementing unipolar current limiting and
  transimpedance amplification. The value of the forward current limit is set by
  $I_\text{bias}$, and the input voltage is approximately 0~V for input
  currents below this limit.
}
\end{figure}

The basic idea behind the
presented circuit design is to use a single bipolar junction transistor (BJT)
$I,V$ characteristic to implement the desired current limiting response while
also providing transimpedance amplification of the DUT current. Packaged
discrete BJTs for radio frequency applications are available with very low
parasitic capacitance, making them highly suitable here for use in the input
stage. The common-base (CB) amplifier configuration is of particular interest as
a high-bandwidth current buffer, featuring a low input impedance and small
feedback capacitance that does not suffer from the Miller effect. With voltage
and current biasing, a CB amplifier can closely approximate the targetted
current limiting $I,V$ characteristic shown in
Fig.~\ref{fig:limiting_characteristic}. A simplified schematic of the input
stage used to accomplish this is shown in Fig.~\ref{fig:simplified_circuit}.

The basic operation of this input stage is straightforward to analyze.
Applying Kirchhoff's current law at the input node, it can be seen that whenever
the DUT current $I_d$ is less than the bias current $I_{\text{bias}}$, the BJT
emitter current $I_E$ is positive and transistor will be in forward-active mode.
In this mode, with an appropriate setting of $V_{\text{bias}} \approx 0.7$~V,
the input voltage $V_{\text{in}}$ will be held close to 0~V due to the high
forward \comment{base-emitter?} transconductance of the BJT. Thus, for either
positive or negative voltages applied to the DUT, the input stage effectively
presents a low impedance to ground as long as $I_d < I_\text{bias}$. As $I_d$
approaches $I_\text{bias}$, the BJT enters cut-off mode, where its effect in the
circuit can be ignored and the input behaves as a current source with $I_d = I_\text{bias}$.

Ideally, the voltage bias $V_\text{bias}$ should be chosen so that the input
current is zero for an input voltage of zero (such that the curve of
Fig.~\ref{fig:limiting_characteristic} intersects the origin). Considering an
approximated Ebers-Moll model \comment{or Shockley diode equation} of the BJT,
it follows that
\begin{equation}\label{eq:1}
V_\text{bias} = -nV_T \log\left(\frac{I_\text{bias}}{I_s} + 1\right),
\end{equation}
where $I_s$ is the saturation current of the base-emitter junction, $V_T\approx
26$~mV is the thermal voltage, and $n$ is the diode ideality factor. The output
of this stage then gives an amplified voltage signal $V_\text{out}$ that is
linearly related to the input current
\begin{equation}\label{eq:2}
  I_d = I_\text{bias} - \left(\frac{1 + \beta}{\beta}\right) \left(\frac{V_{CC} - V_\text{out}}{R_C}\right),
\end{equation}
where $\beta$ is the forward common-emitter current gain of the
NPN transistor.

\begin{figure}[h!]
\setlength{\fboxrule}{0pt}%
\setlength{\fboxrule}{0pt}%
\centering
\fbox{
\maxsizebox{\columnwidth}{!}{\begin{tikzpicture}

\def \BJTscale{1.5};			
\def \offsetUp{1.5};			
\def \offsetDown{0};		
\def \offsetDiffMeas{.3};		

\node (probe) at (0,0) {};
\node[circ]  at (probe) {};

\node[jump crossing,right = 1.5cm of probe] (mirror) {};


\node[left of=probe, node distance=.5cm](needle){};
\node[above of = needle, node distance=0.3cm,anchor=east,xshift=8pt]{Contact Probe};
\draw (needle.center)--++(0,0.05)--++(-.8,0)--++(-1,-0.4)--++(1,.3)-|(needle.center);
\draw (probe.center)--(needle.center);

\draw (probe.center) to[short] ++(0,\offsetUp) node[npn, anchor=E, scale=\BJTscale](Q1){};
\draw (Q1.C) to[short] ++(0,\offsetDiffMeas)node[circ](Vprobe){} to[R, name=R1,l2 = $\mathrm{R_1}$ and $\mathrm{2 k\Omega}$, l2 valign=t, l2 halign=r ] ++(0,1.5) node(leftVcc){};

\draw (probe.center) to[short] ++(0,\offsetDown) node[npn, anchor=C, scale=\BJTscale](Q2){};
\draw (Q2.E) node[npn, anchor=C, scale=\BJTscale](Q3){};
\draw (Q3.E) to[R,l2_ = $\mathrm{R_2}$ and $\mathrm{2 k\Omega}$, l2 valign=t, l2 halign=r] ++(0,-1)--++(0,-.5) node[](leftVee){};

\draw (mirror.center) to[short] ++(0,\offsetUp) node[npn, anchor=E, xscale=-1, scale=\BJTscale](Q4){};
\draw (Q4.C) to[short] ++(0,\offsetDiffMeas) node[jump crossing](Vmirror){} to[R, name=R2,l2=$\mathrm{R_3}$ and $\mathrm{2 k\Omega}$, l2 valign=t, l2 halign=r] ++(0,1.5) node(rightVcc){};
\draw (Vprobe) -- (Vmirror.west);

\draw (mirror.center) to[short] ++(0,\offsetDown) node[npn, anchor=C, xscale=-1, scale=\BJTscale](Q5){};
\draw (Q5.E) node[npn, anchor=C, xscale=-1, scale=\BJTscale](Q6){};
\draw (Q6.E) to[R,l2_ = $\mathrm{R_4}$ and $\mathrm{2 k\Omega}$, l2 valign=t, l2 halign=r] ++(0,-1) --++(0,-.5)node[](rightVee){};

\node at (Q1.center){Q$_1$};
\node at (Q2.center){Q$_2$};
\node at (Q3.center){Q$_3$};
\node at (Q4.center){Q$_4$};
\node at (Q5.center){Q$_5$};
\node at (Q6.center){Q$_6$};

\node[circ] (Vcc) at ($(leftVcc)!0.5!(rightVcc)$) {}; 			
\draw (leftVcc.center) -- (rightVcc.center);
\draw (Vcc) to[short] ++(0,.1) node[vcc]{+6V};

\node[circ] (Vee) at ($(leftVee)!0.5!(rightVee)$) {}; 			
\draw (leftVee.center) -- (rightVee.center);
\draw (Vee) to[short] ++(0,-.1) node[vee]{-10V};

\draw (Q2.B) -- ++(0,-.5)node[sground,scale=1]{};
\draw (Q5.B) -- ++(0,-.5)node[sground,scale=1]{};

\draw (Q3.B) to[short, *-o] ++(-.5,0) node[left]{$\mathrm{V_{c}}$};
\node[jump crossing] (Q6jump) at ($(Q6.E) + (0,.3)$){};
\node[jump crossing] (Q3jump) at ($(Q3.E) + (0,.3)$){};
\draw (Q6.B) |- (Q6jump.east);
\draw (Q6jump.west) -- (Q3jump.east);
\draw (Q3jump.west) -|(Q3.B);

\node[below = .2cm of Q4.E](FBpoint){};
\node[circ] at (FBpoint){};
\draw (FBpoint.center) -- ++(2,0) node[op amp, noinv input up, anchor=-, scale=.9](biasop){OPA277};
\draw (biasop.+)  node[sground, scale=1]{};

\node[jump crossing] (Q4jump) at ($(Q4.E) + (0,.3)$){};
\node[jump crossing] (Q1jump) at ($(Q1.E) + (0,.3)$){};
\draw (biasop.out) |- (Q4.B) node[circ]{}  |- (Q4jump.east);
\draw (Q4jump.west) -- (Q1jump.east);
\draw (Q1jump.west) -|(Q1.B);

\node[right = 1.88 of mirror.center](posVop) {};
\node[op amp, noinv input up ,anchor=+, scale=.9] (Vop) at (posVop) {THS3091};
\draw (probe.center) -- (mirror.west);
\draw (mirror.east) -- (Vop.+);
\draw (Vop.-) -- ++(0,-1) node[circ](Vfb){} to[R,l=$\mathrm{R_{9}}$,a=$1.21\mathrm{k}\Omega$, l2 valign =b] (Vfb-| Vop.out) -- (Vop.out) node[circ]{};
\draw (Vfb) --++(0,-.2) to[R,l2=$\mathrm{R_{8}}$ and $1.21\mathrm{k}\Omega$, l2 halign=l, l2 valign =t] ++(0,-2) node[sground, scale=1]{};

\draw (Vop.out) to[R,l=$\mathrm{R_{10}}$,a=$50\Omega$,-o] ++(2,0) node[right]{$\mathrm{V_{out}}$};

\node [muxdemux, muxdemux def={NL=4, NR=1, NT=0, NB=0, w=4,  Lh=5, Rh=0}, 
anchor=lpin 3](AD8130) at ($(Vmirror)+(2,0)$) {AD8130};
\node[right=.2] at (AD8130.lpin 3) {$+$};
\node[right=.2] at (AD8130.lpin 4) {$-$};
\node[right=.2, font=\tiny] at (AD8130.lpin 2) {REF};
\node[right=.2, font=\tiny] at (AD8130.lpin 1) {FB};

\draw (Vmirror.east) -- (AD8130.lpin 3);
\draw (AD8130.lpin 4) -- (AD8130.lpin 4 -| Vmirror) node[circ]{};

\draw (AD8130.lpin 1) -- ++(0,.6) node[circ](fb){} to[R,l=$\mathrm{R_{6}}$,a=$499\Omega$] (fb -| AD8130.rpin 1) --(AD8130.rpin 1) node[circ]{}to[R,l=$\mathrm{R_{7}}$,a=$50\Omega$,-o] ++(1.5,0) node[right]{$\mathrm{I_{out}}$};
\draw (AD8130.lpin 2) node[sground, scale=1]{} ;
\draw (fb) to[R,a=$\mathrm{R_{5}}$,l=$499\Omega$] ++(-1.5,0) node[sground, scale=1]{} ;

\end{tikzpicture}}
      }
\caption{\label{fig:full_circuit}
  Full schematic for the current limiting probing circuit. All BJT devices are
ON Semiconductor part no. NSVF5501SKT3G. Transistors Q1-Q3 perform the current limiting
function, with the current limit controlled by the input signal $V_{c}$.
Regulated power supplies providing $\pm$10~V and +6~V are not shown.
}
\end{figure}
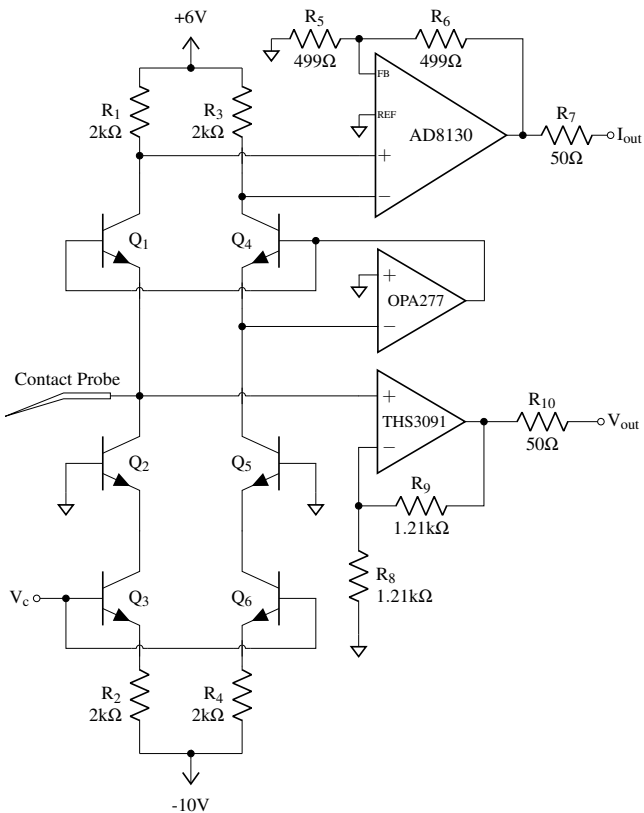

\begin{figure}[h!]
{
\setlength{\fboxsep}{0pt}%
\setlength{\fboxrule}{1pt}%
\fbox{\includegraphics[width=\columnwidth]{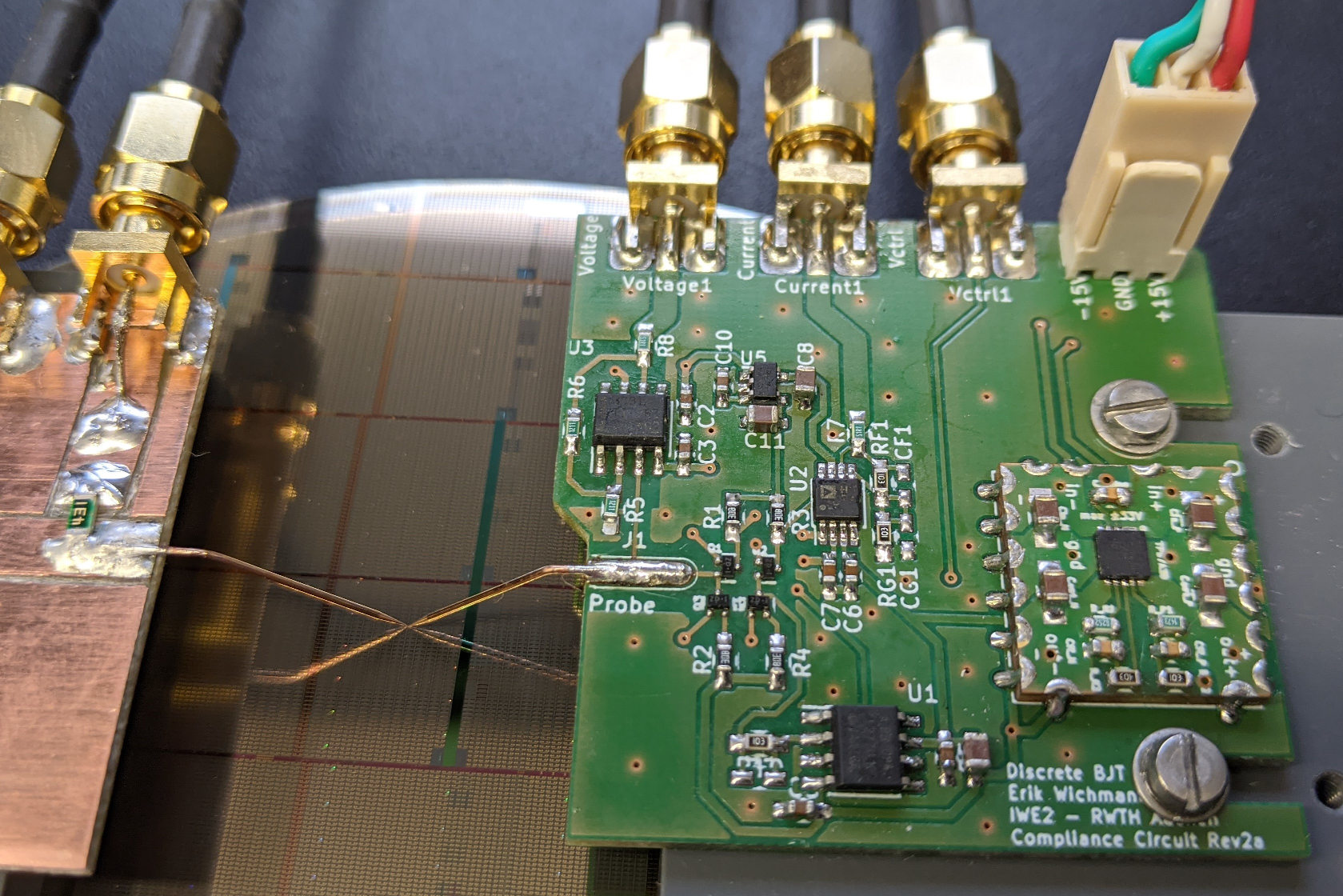}}
}
\caption{\label{fig:pcb_photo}
  Photograph of the probing circuit board contacting a prototype RRAM device. Left and right probes are
  mounted on independent micropositioners.
}
\end{figure}

A full circuit diagram expanding on this concept is given in
Fig.~\ref{fig:full_circuit}, with a prototype PCB layout also pictured in
Fig.~\ref{fig:pcb_photo}. Here, Q$_1$ is the CB amplifier corresponding to that
depicted in Fig.~\ref{fig:simplified_circuit}, and a nearly ideal voltage
controlled current source is realized by the emitter degenerated cascode
amplifier formed by Q$_2$, Q$_3$, and R$_2$. The dependence of the current limit
$I_\text{bias}$ on the control voltage $V_c$, which is approximately linear for
$I_\text{bias} > 100~\mu\text{A}$, is calibrated for $V_c$ values between
$-10~\text{V}$ and $-1~\text{V}$ by an SMU measurement. The $V_c$ signal is then
generated according to interpolation of the calibration table at the desired
$I_\text{bias}$ values.

Further circuitry in Fig.~\ref{fig:full_circuit} is included to null voltage
offsets and condition the output signals for transmission to 50~$\Omega$
oscilloscope inputs. From Eq.~\ref{eq:1}, it is seen that the ideal value of
$V_\text{bias}$ depends slightly on the value of $I_\text{bias}$. Therefore,
simply using a constant value of $V_\text{bias}$ would create offset voltages at
the input terminal on the order of $10-100$~mV as $I_\text{bias}$ is varied. To
automatically compensate this effect, a reference path $R_3, Q_4, Q_5, Q_6, R_4$
mirrors the components $R_1, Q_1, Q_2, Q_3, R_2$, and is used to actively zero
the input offset for all values of $I_\text{bias}$ via OPA277. This same
structure also generates a reference voltage for a differential measurement
performed by AD8130, producing a low-offset output signal $I_\text{out}$
proportional to the input current. A voltage follower (THS3091) with very low
input capacitance (0.1~pF) is also placed directly at the input node, providing
a simultaneous measurement of the DUT voltage drop.

\section{\label{sec:results} Measurement Results}

Current overshoots accompanying sudden negative resistance transitions are
suppressed in our measurement scheme by minimizing the capacitance at the input
node of the CLA. This is done by careful selection of the input transistors and
by avoiding proximity of input traces to the ground plane. However, the
parasitic capacitance cannot be fully eliminated and the potential to
create overshoots inevitably remains. Since overshoot transients tend to play a
critical role in switching behavior in practice, it is important to characterize
and model them.

In general, the time-dependent $I,V$ trajectory of a current overshoot is not
solely a characteristic of the measurement setup, but is determined by the
coupled dynamics of the DUT conductance and the driving circuitry. The duration
and amplitude of the overshoot therefore depends on the type and history of the
RS cell being measured, and is not easily reproducible. To measure the
overshoot characteristic in a standardized way, a test sample designed to
imitate \comment{mimic} the resistive switching action was constructed using
surface mount components. A mechanical reed relay in series with a 1.2~k$\Omega$
and in parallel with 100~k$\Omega$ was found to be well suited for this purpose,
providing a controllable sub-nanosecond transition between two discrete
resistance levels with negligible parasitic effects.

\begin{figure}[h!]
\includegraphics[width=\columnwidth]{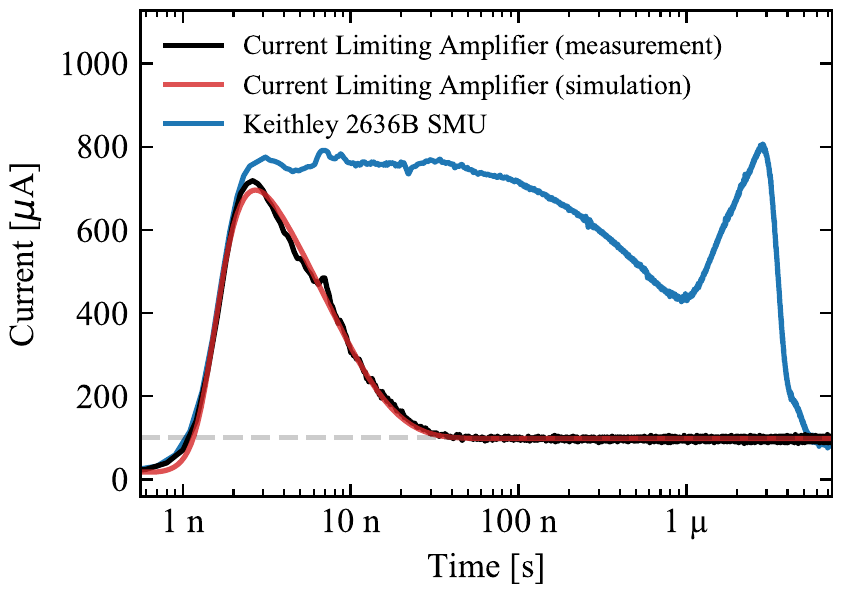}
\caption{\label{fig:overshoot_sim} Current overshoot characterization using a reed
relay to abruptly switch from 100~k$\Omega$ to 1.2~k$\Omega$ at time 1~ns with 1~V
applied and with a current limit of 100~$\mu$A. Under these conditions, the CLA
returned to the current limit in $\sim$20~ns, whereas a commercial SMU produced a
more complex overshoot response lasting several microseconds.}
\end{figure}

With the reed switch connected in the position of the DUT and biased by 1~V, the
current transient following a resistance transition was measured with 350~MHz
bandwidth (Fig.~\ref{fig:overshoot_sim}). Close agreement of the transient was
found with the solution of a differential equation describing the charging of
the CLA input node,
\begin{equation}\label{eq:3}
C_p \frac{dV_d}{dt} = I_\text{bias}\left[1 - \exp{\left( \frac{V_d - V_a}{V_T} \right)}\right] - \frac{V_d}{R_d},
\end{equation}
where $C_p=5.7$~pF is the parasitic capacitance at the input, $V_d$ is the DUT
voltage drop, $V_a$ is the applied voltage, and $R_d$ is the DUT resistance
(here assumed a step function in time). Note that $C_p$ includes the
self-capacitance of the measured cell, which is approximately 0.5~pF for the
reed relay circuit. This should be taken into consideration in the memory cell
design itself, where thin dielectric layers and large contact pads or device
areas can contribute significantly to the total $C_p$, which intrinsically
degrades the overshoot performance. Given the single parameter $C_p$, the simple
model of Eq.~\ref{eq:3} is expected to accurately characterize the transient
response of the CLA circuit, and should be incorporated with a physical device
model to properly understand the complete picture of the coupled system during a
measurement.


For comparison, the current overshoot transient induced using a modern SMU was
measured under identical conditions. For the first $1~\mu$s after the resistance
transition, the transient begins with the discharge of a 1~m coaxial cable which
must be used to connect the instrument. Between $1-10~\mu$s, a proprietary
feedback circuit is engaged and produces a long unpredictable current excursion
before undershooting and eventually settling to the programmed current
compliance level. Relative to this, the overshoot duration is reduced in the CLA
measurement by over two orders of magnitude.

\begin{figure}[h]
\includegraphics[width=\columnwidth]{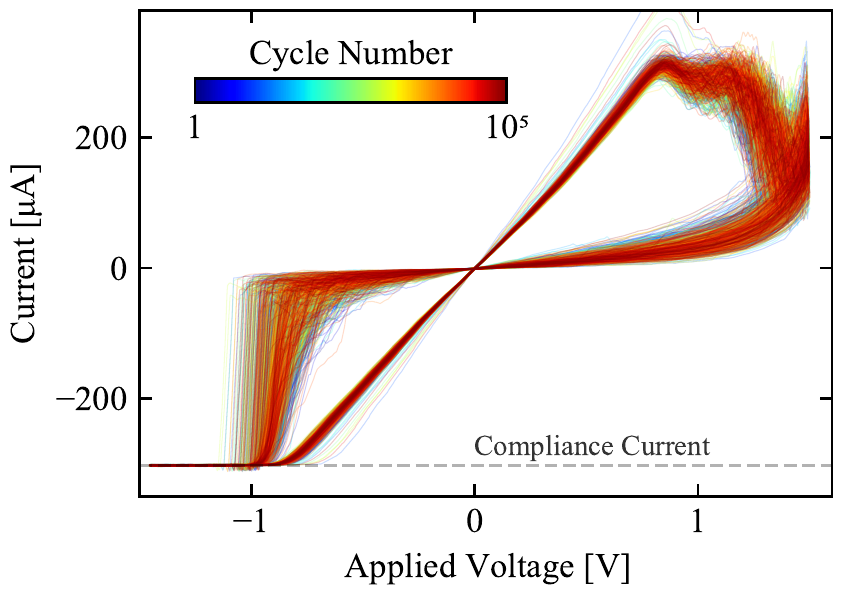}
\caption{\label{fig:mass_ivloops} A measurement of $10^5$ consecutive $I,V$
loops collected in one second with the CLA circuit using a triangular voltage
excitation and 300~$\mu$A current limit. Data is smoothed by a 15 sample moving
average, and every 100th cycle is plotted. To conform to plotting convention, the
applied voltage is defined as the negative of the AWG voltage.
}
\end{figure}

To demonstrate the RS cycling operation using the external CLA circuit, we
tested a TaOx-based nano-scaled (100~nm) RRAM device of a design which was known
not to survive repeated switching using a commercial SMU. With the CLA input
connected to the DUT top electrode, the current limit was set to 300~$\mu$A and
a triangular voltage signal with period 10~$\mu$s and amplitude 1.5~V was
applied to the DUT bottom electrode using a Rigol DG5102 AWG. The applied
voltage and device current were sampled at 1.25 GS/s using a Picoscope 6404D
deep-storage oscilloscope. In a single measurement lasting only one second, $10^5$
full $I,V$ loops were successfully collected, each containing 1,564 8-bit $I,V$
samples (Fig.~\ref{fig:mass_ivloops}). It is furthermore possible to collect
millions of such cycles in a practical amount of time by collating multiple
measurement shots, creating powerful datasets for statistical evaluation of RS
devices.

\section{\label{sec:conclusion}Conclusion}
Resistive switching devices are promising building blocks for future memory and
neuromorphic architectures, with the salient property of large cycle-to-cycle
variability. Conventional lab measurements of these cells commonly represent
very different conditions from integrated systems, and often have unclear
implications for device applications. In particular, current overshoots during
runaway resistance transitions hinder the ability to control and characterize
the switching process. In this work, an external current limiting amplifier was
developed to reduce the overshoot effect and allow for measurements of full
$I,V$ loops at $\sim 10^6$ times faster rates than commercial SMUs. The minimal
design with low transistor count is relatively robust against load-induced
instability and has the important advantage that its response is accurately
predictable using a few idealized component models.

\comment{
\begin{acknowledgments}
  We acknowledge the contributions of people and institutions.
\end{acknowledgments}
}

\bibliography{compliancepaper}

\end{document}